\documentclass[12pt,preprint]{aastex}

\def\ni{\noindent}

\def\cm{{\rm\,cm}}

\def\gm{{\rm\,g}}

\def\yr{{\rm\,yr}}

\def\AU{{\rm\,AU}}

\def\pomega{\tilde{\omega}}

\begin{document}

\shortauthors{Chiang et al.}
\shorttitle{The 5:2 and Trojan Resonances}

\title{Resonance Occupation in the Kuiper Belt: Case Examples of the 5:2 and
Trojan Resonances}

\author{E.~I.~Chiang\altaffilmark{1}, A.~B.~Jordan\altaffilmark{1},
R.~L.~Millis\altaffilmark{2}, M.~W.~Buie\altaffilmark{2},
L.~H.~Wasserman\altaffilmark{2}, J.~L.~Elliot\altaffilmark{2,3,4},
S.~D.~Kern\altaffilmark{3}, D.~E.~Trilling\altaffilmark{5},
K.~J.~Meech\altaffilmark{6}, \& R.~M.~Wagner\altaffilmark{7}}

\altaffiltext{1}{Center for Integrative Planetary Sciences,
Astronomy Department, University of California at Berkeley,
Berkeley, CA~94720}
\altaffiltext{2}{Lowell Observatory, 1400 West Mars Hill Road, Flagstaff,
AZ~86001}
\altaffiltext{3}{Department of Earth, Atmospheric, and Planetary Sciences,
Massachusetts Institute of Technology, 77 Massachusetts Avenue, Cambridge,
MA~02139}
\altaffiltext{4}{Department of Physics, Massachusetts Institute of Technology,
77 Massachusetts Avenue, Cambridge, MA~02139}
\altaffiltext{5}{University of Pennsylvania, Department of Physics and
Astronomy, David Rittenhouse Laboratory, 209 S. 33rd St., Philadelphia,
PA~19104}
\altaffiltext{6}{Institute for Astronomy, 2680 Woodlawn Drive, Honolulu,
HI~96822}
\altaffiltext{7}{Large Binocular Telescope Observatory, University of Arizona,
Tucson, AZ~85721}
\email{echiang@astron.berkeley.edu}

\begin{abstract}
As part of our ongoing Deep Ecliptic Survey (DES) of the Kuiper belt,
we report on the occupation of the 1:1 (Trojan), 4:3, 3:2, 7:4, 2:1, and 5:2
Neptunian mean-motion resonances (MMRs).
The previously unrecognized occupation of
the 1:1 and 5:2 MMRs is not easily
understood within the standard model of resonance sweeping
by a migratory Neptune over an initially dynamically cold belt.
Among all resonant Kuiper belt objects (KBOs),
the three observed members of the 5:2 MMR discovered
by DES possess the largest semi-major axes ($a \approx 55.4\AU$),
the highest eccentricities ($e \approx 0.4$), and substantial
orbital inclinations ($i\approx 10\degr$). Objects (38084) 1999HB$_{12}$
and possibly 2001KC$_{77}$ can librate with modest amplitudes
of $\sim$$90\degr$ within the 5:2 MMR for at least 1 Gyr. Their trajectories
cannot be explained by close encounters with Neptune
alone, given the latter's current orbit. The dynamically hot orbits of
such 5:2 resonant KBOs, unlike hot orbits of previously
known resonant KBOs, may imply that these objects were pre-heated
to large inclination and large eccentricity prior to resonance capture by
a migratory Neptune. Our first discovered
Neptunian Trojan, 2001QR$_{322}$, may not owe its existence
to Neptune's migration at all. The trajectory of 2001QR$_{322}$ is remarkably
stable; the object can undergo tadpole-type libration about
Neptune's leading Lagrange (L4) point for at least 1 Gyr
with a libration amplitude of $24\degr$.
Trojan capture probably occurred while Neptune accreted
the bulk of its mass. For an assumed albedo of 12--4\%,
our Trojan is $\sim$130--230 km in diameter.
Model-dependent estimates place the total number
of Neptune Trojans resembling 2001QR$_{322}$
at $\sim$20--60. Their existence might rule out violent
orbital histories for Neptune.
\end{abstract}


\section{INTRODUCTION}
\label{intro}
A fraction of Kuiper belt objects (KBOs) occupy low-order, exterior
mean-motion resonances (MMR) established by Neptune.
Among the most well-known resonant KBOs are the Plutinos
which occupy the 3:2 MMR (Jewitt \& Luu 2000).
Plutinos have substantial orbital eccentricities,
$0.1 \lesssim e \lesssim 0.3$, an observation
commonly interpreted to imply that Neptune migrated
outwards by several AU early in the history of the solar system
(Malhotra 1995). The standard model of resonant capture and adiabatic
excitation by a migratory Neptune predicts the 2:1, 5:3, 7:4, 3:2, and
4:3 MMRs to be occupied by high eccentricity objects (Malhotra et al.~2000;
Chiang \& Jordan 2002, hereafter CJ). Occupation of the 4:3 MMR
and possibly of the 2:1 MMR has been reported by Nesvorny \& Roig (2001).
Implications of Neptune's migration for the Plutinos
and the ``Twotinos'' (2:1 resonant KBOs) are explored in detail by CJ.

We report here, as part of the ongoing survey of the Kuiper Belt
by the Deep Ecliptic Survey Team (Millis et al.~2002; Elliot et al.~2003),
the previously unrecognized occupation of
the 5:2 and 1:1 (Trojan) Neptunian resonances.
The three observed members of the 5:2 MMR stand out among all resonant
KBOs in having the largest semi-major axes
($a \approx 55.4\AU$), the highest eccentricities
($e \approx 0.4$), and substantial orbital inclinations ($i\approx 10\degr$).
We will see that their dynamically hot orbits
cannot be interpreted as initially dynamically cold orbits
that were modified purely by resonance sweeping.
Their existence points to another dynamical excitation mechanism
that likely operated prior to Neptune's migration.

Our discovery of the first Neptunian Trojan librating about
the leading Lagrange (L4) point of Neptune vindicates
theoretical suggestions as to the long-term orbital stability
of Neptunian Trojans (Holman \& Wisdom 1993; Holman 1995; Gomes 1998; Nesvorny
\& Dones 2002). For example, Nesvorny \& Dones (2002) find
that about 50\% of their hypothesized Neptunian Trojan population survives
for 4 Gyr despite perturbations exerted by the other giant planets.
The stability of Neptune's Trojan population contrasts
with the instability characterizing Saturnian and Uranian Trojans
on $10^8\yr$ timescales (Nesvorny \& Dones 2002; Gomes 1998).

In \S\ref{resmem}, we outline our procedure
for identifying resonant KBOs in the face of
observational uncertainties in their orbits,
and describe the dynamical characteristics of
our 5:2 and 1:1 resonant KBOs.
Results of Gyr-long orbit integrations of our Trojan
are presented at the end of this section.
In \S\ref{models}, we briefly assess the plausibility
of some theoretical scenarios that attempt to explain the
observed pattern of resonance occupation.
We consider models in which Neptune either sweeps objects
into its resonances by virtue of its migration, or
populates the resonances by direct, violent gravitational scattering.
A summary of our results, interpreted in the context of theoretical models,
is provided in \S\ref{sum}.

\section{Observed Resonance Membership}
\label{resmem}

\subsection{Classification Procedure}
\label{clasproc}
An object occupies a MMR if the resonant argument
associated with that MMR librates. For the
5:2, $e^3$ (third degree in the eccentricity of the KBO)
Neptunian MMR, the argument
is $\phi_{5:2} = 5\lambda - 2\lambda_N - 3\pomega$,
where $\lambda$ and $\pomega$ are the mean
longitude and longitude of pericenter
of the object, respectively, and $\lambda_N$ is the mean
longitude of Neptune. For the 1:1, $e^0$ Neptunian MMR,
the argument equals $\phi_{1:1} = \lambda - \lambda_N$.

Testing for libration is a straightforward matter
of integrating forward (or backward) the trajectory
of an object in the gravitational fields of the
Sun and the planets. A secure identification
of a resonant KBO is made difficult by often substantial
uncertainties in the initial position and velocity
of the object, i.e., uncertainties in the osculating
Keplerian ellipse fitted to astrometric observations.
Bernstein \& Khushalani (2000) derive a formalism
for estimating these errors that is tailored
for short arc astrometric observations. Our Deep
Ecliptic Survey (DES; Millis et al.~2002; Elliot et al.~2003)
utilizes their formalism.

\placefigure{fig1}
\begin{figure}
\epsscale{0.55}
\vspace{-1in}
\plotone{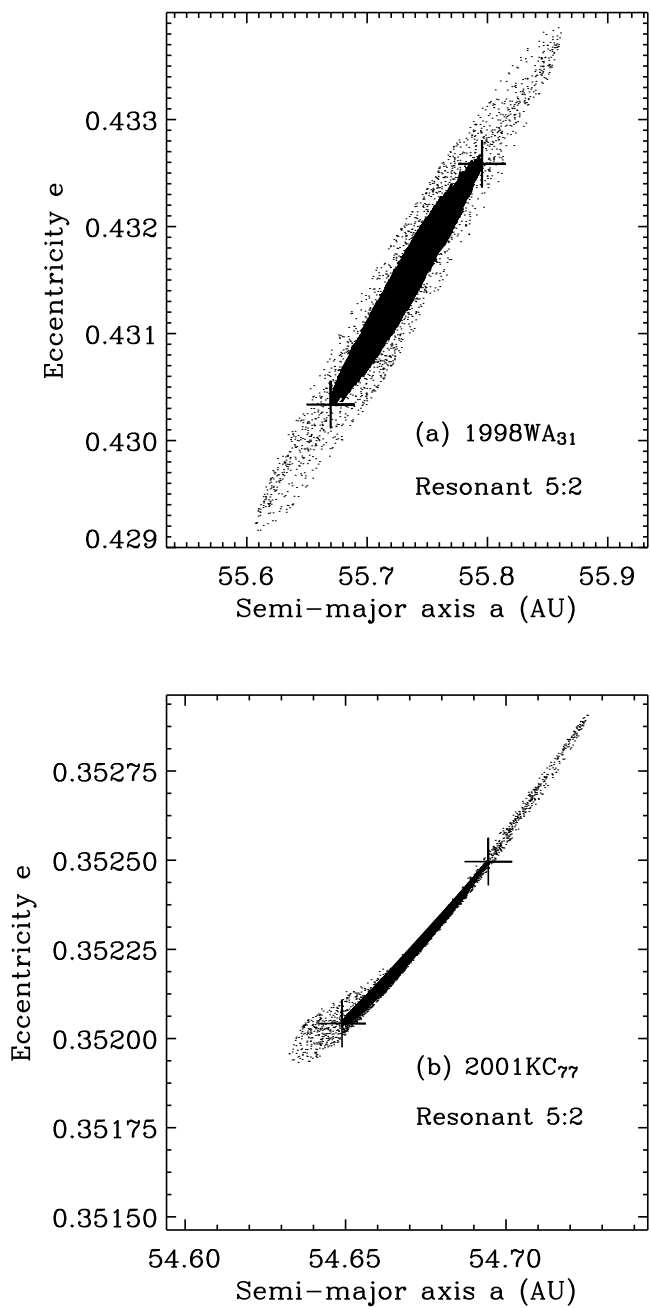}
\caption{Projections onto the $a$-$e$ plane of the
$1\sigma$ and $2\sigma$ confidence surfaces
in the 6-dimensional phase space of possible
osculating orbits for 5:2 resonant KBOs (a) 1998WA$_{31}$,
and (b) 2001KC$_{77}$. Solid black areas correspond
to $1\sigma$ confidence regions, while speckled
areas correspond to $2\sigma$ confidence regions.
The exact center of each plot corresponds to orbit
solution 1, while crosses denote orbit solutions 2 and 3.}
\label{aecon}
\end{figure}

\placefigure{fig2}
\begin{figure}
\epsscale{0.55}
\vspace{-1in}
\plotone{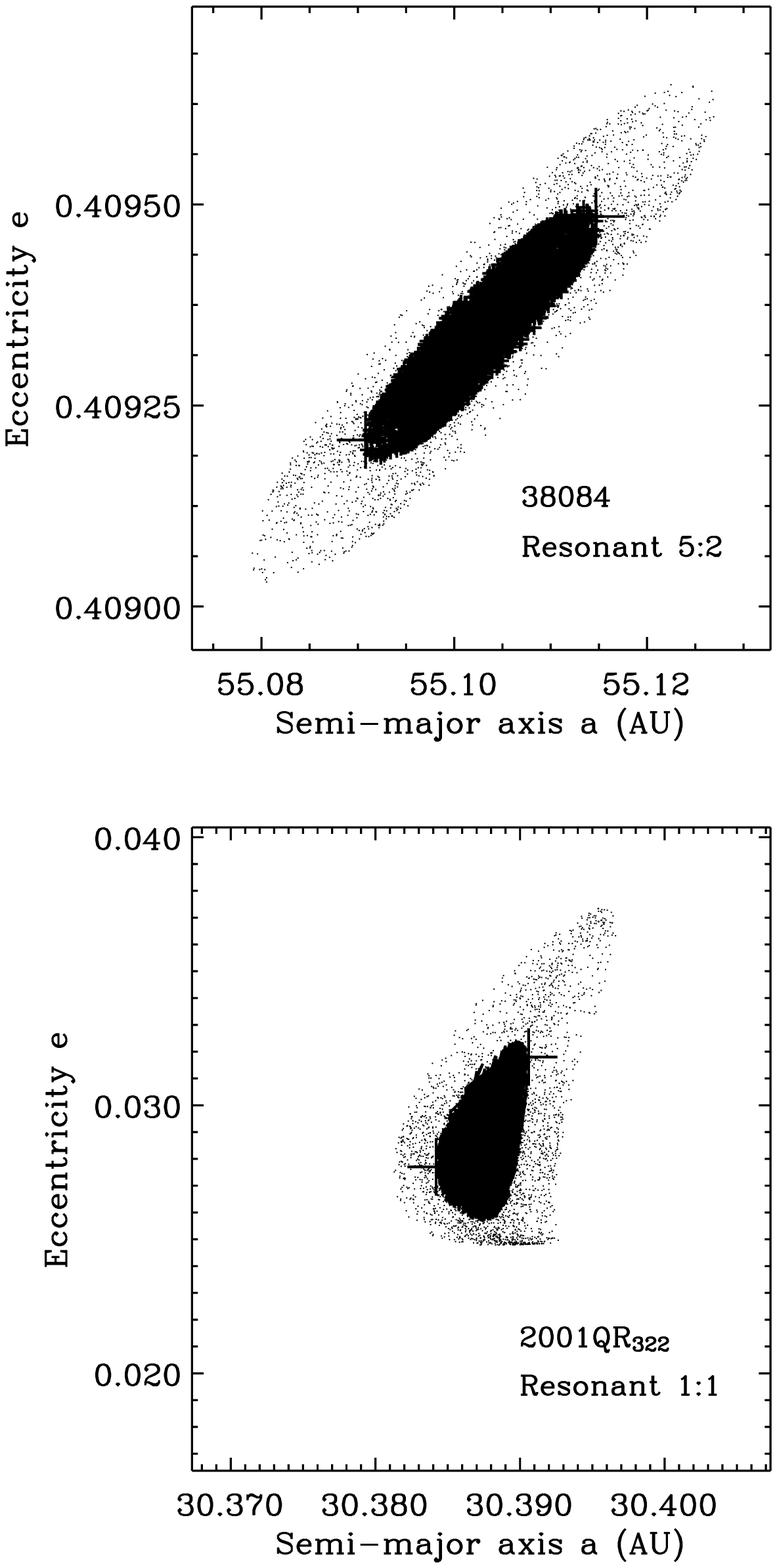}
\caption{Projections onto the $a$-$e$ plane of the
$1\sigma$ and $2\sigma$ confidence surfaces
in the 6-dimensional phase space of possible
osculating orbits for
(a) (38084) 1999HB$_{12}$, a 5:2 resonant KBO, and
(b) 2001QR$_{322}$, a 1:1 resonant KBO. Solid black areas correspond
to $1\sigma$ confidence regions, while speckled
areas correspond to $2\sigma$ confidence regions.
The exact center of each plot corresponds to orbit
solution 1, while crosses denote orbit solutions 2 and 3.}
\label{aecon2}
\end{figure}

Figures \ref{aecon} and \ref{aecon2} depict the 1$\sigma$ and 2$\sigma$
confidence regions projected onto the $a$--$e$ plane
of our three 5:2 resonant candidates and our one
1:1 resonant candidate.
Uncertainties in $a$ and $e$ for these
particular objects are small, of order 0.1\%,
thanks to the relatively extended, 1+ year-long arcs
of astrometry available for these KBOs.
All elements reported in this paper are osculating, heliocentric
elements referred to the J2000 ecliptic plane, evaluated at epoch 2451545.0 JD.

Surveying for libration in the 6-dimensional
confidence volume of possible initial orbits
for each of hundreds of KBOs discovered by our Deep Ecliptic
Survey is daunting. We proceed with a more
limited agenda; in the 6-D error volume
appropriate to a given KBO, we integrate, in addition
to the nominal best-fit (initial) osculating orbit,
two other solutions that lie on the 1$\sigma$
confidence surface and that have the greatest and least semi-major
axes. We refer to these sets of initial conditions as orbit
solutions 1, 2, and 3, respectively. The other 5 orbital
elements are adjusted according to their correlation with
semi-major axis on the 1$\sigma$ confidence surface.
We favor exploring the widest
excursion in semi-major axis because that is the parameter
that most influences resonance membership.
The next most important parameter is eccentricity;
however, as is evident in Figures \ref{aecon} and \ref{aecon2},
deviations in $a$ and $e$ are often strongly
correlated, so that exploring the greatest
deviation in $a$ often implies
that we are also exploring the greatest deviation
in $e$. Our choice of focussing on variations in $a$
is further supported by the fact that fractional errors
in $a$ (and $e$) are slower to converge
to zero than errors in $i$ (Millis et al.~2002).

We numerically integrate three sets of initial
conditions for each of 204 KBOs
discovered by the DES collaboration as of April 9 2002 and
given preliminary designations by the Minor Planet Center. We employ
the regularized, mixed variable symplectic
integrator, swift\_rmvs3, developed by
Levison \& Duncan (1994) and based
on the N-body map of Wisdom \& Holman (1991).
We include the influence of the four giant
planets, treat each KBO as a massless test
particle, and integrate trajectories
forward for $3 \times 10^6\yr$ using a timestep
of 50 days, starting at Julian date 2451545.0.
Initial positions and velocities for all objects are computed
using the formalism of Bernstein \& Khushalani (2000)
in the case of short-arc orbits, and from E.~Bowell's
database in the case of long-arc orbits (see Millis et al.~2002).
The relative energy error over the integration is
bounded to less than $10^{-7}$. One hundred and seven
different resonant arguments are examined for
libration. Full details of our procedure
are provided in Elliot et al.~(2003).

\placefigure{fig2}
\begin{figure}
\epsscale{0.85}
\vspace{-0.5in}
\plotone{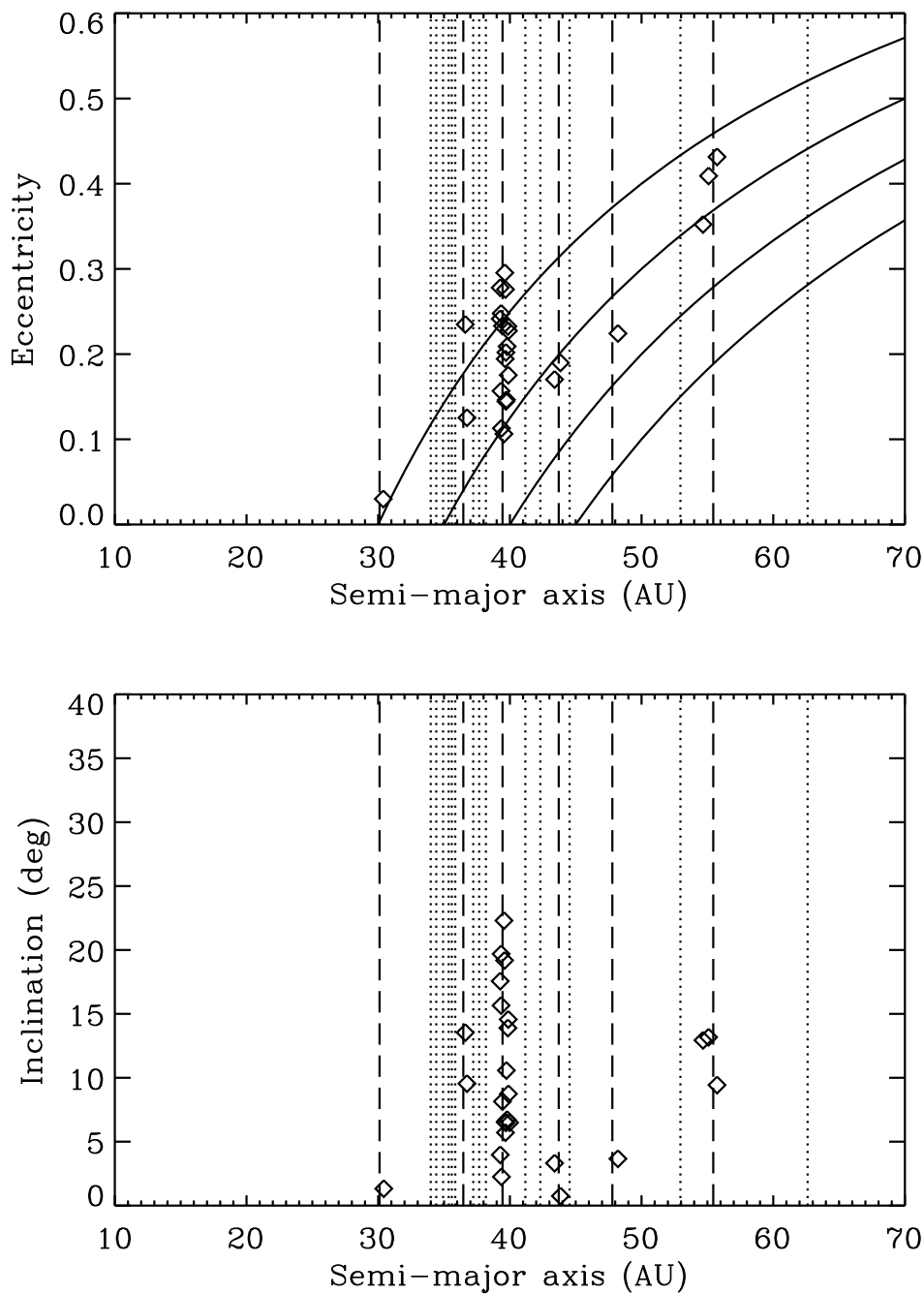}
\caption{Eccentricities, inclinations, and semi-major axes of resonant KBOs
found in the Deep Ecliptic Survey. For all displayed objects, fractional
$1\sigma$ uncertainties in semi-major axis range from 0.003\% to 3.5\%, and
orbit solutions 1, 2, and 3 yield consistent orbital classifications. Open
diamonds represent resonant objects only; non-resonant objects will be
presented
by Elliot et al.~(2003). Vertical lines indicate locations of nominal resonance
with Neptune; dotted lines indicate uninhabited resonances, while dashed lines
indicate inhabited resonances. Solid curves correspond to perihelion distances
of 30, 35, 40, and 45 AU. Resonances with secure members include, in order of
increasing distance from Neptune, the 1:1, 4:3, 3:2, 7:4, 2:1, and 5:2 MMRs.
}
\label{ae}
\end{figure}

Some results of this procedure are showcased
in Figure \ref{ae}, which contains only a small subset
of the data to be released by Elliot et al.~(2003).
Only ``secure'' resonant objects are displayed; by ``secure,''
we mean that the
1$\sigma$ fractional uncertainties in semi-major axis
are less than 10\% and that all
three sets of initial conditions give consistent
orbit classifications over 3 Myr.
A resonant object is one for which all three
orbit solutions yield libration of one or more of
the same resonant arguments; non-resonant objects
exhibit no libration of any resonant argument among the three solutions.
The locations of the points in Figure \ref{ae} correspond
to the semi-major axes, eccentricities, and inclinations
at the start of the integration. Error
contours are much smaller than the sizes
of the symbols in most cases. Dashed lines
delineate the locations of nominal resonance
with Neptune. In addition to confirmed
librators in the 4:3, 3:2, 7:4, and 2:1 resonances,
the 1:1 and 5:2 resonances contain one and three members,
respectively.

Figure \ref{ae} displays only objects
discovered by the DES collaboration. Other groups
have reported occupation of the 3:2, 4:3, and 2:1 resonances.
For example, Nesvorny \& Roig (2001)
have reported KBOs occupying the 4:3 MMR and possibly the
2:1 MMR. When we integrate the trajectories of non-DES objects,
we confirm the results of Nesvorny \& Roig (2001)
that 1995DA$_2$ inhabits the 4:3 MMR and that (20161) 1996TR$_{66}$
and 1997SZ$_{10}$ inhabit the 2:1 MMR. The names of
resonant objects discovered by the DES team
and by non-DES teams are contained in Table 2.

Whereas the 4:3, 3:2, 7:4, and 2:1 are predicted
by the standard migration model for Neptune
to be substantially populated
(see Figures 3 and 4 of CJ), the 5:2 and 1:1 resonances
are not. We focus our attention now on the newly discovered
members of the 5:2 and 1:1 MMRs, to investigate the constraints
they place on the dynamical history of the Kuiper belt.

\subsection{Observed Members of the 5:2 MMR}
\label{fivetwo}

\begin{figure}
\epsscale{0.85}
\vspace{-1in}
\plotone{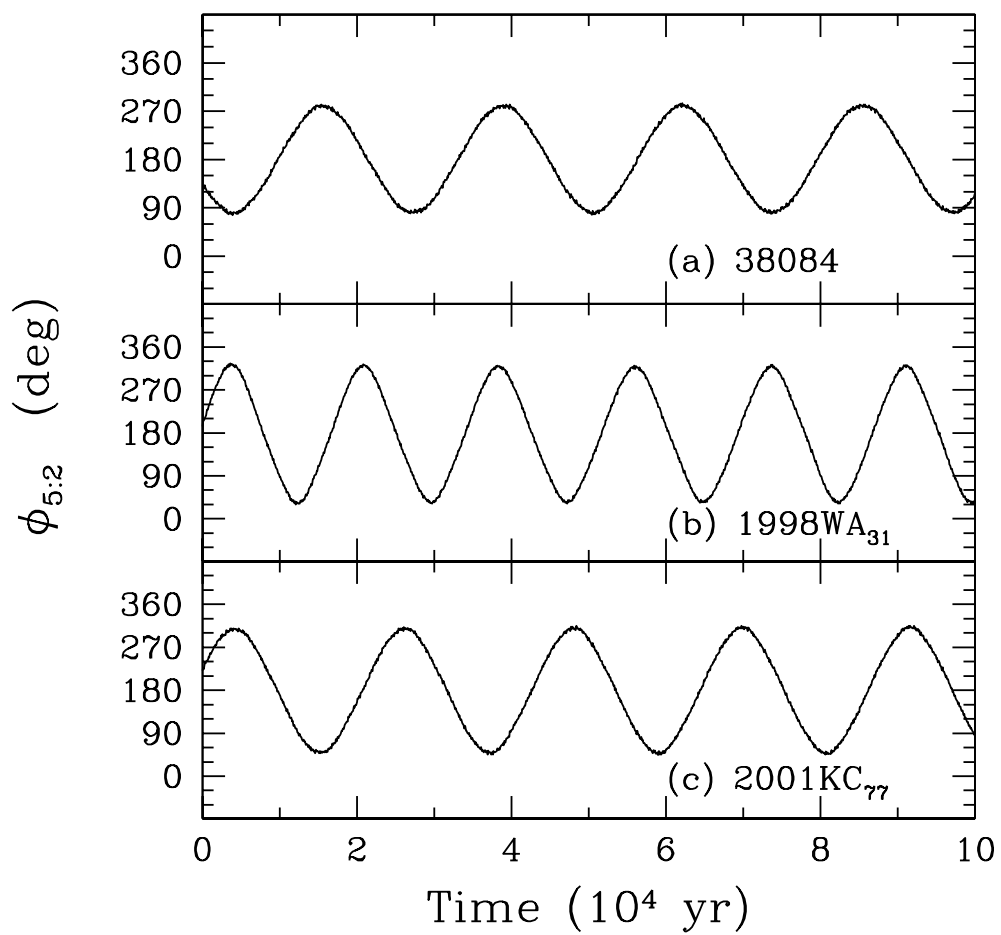}
\caption{Libration of the resonant argument, $\phi_{5:2}$,
for our observed members of the 5:2 resonance. Integrations
begin with nominal best-fit initial conditions (orbit solution 1).
}
\label{wavy}
\end{figure}

Evolutions of the resonant argument, $\phi_{5:2}$,
for objects 1998WA$_{31}$, (38084) 1999HB$_{12}$, and 2001KC$_{77}$
are displayed in Figure \ref{wavy}.
The integrations shown begin with nominal
best-fit initial conditions; the other two sets
of initial conditions yield nearly identical results.
The resonant argument, $\phi_{5:2}$, librates in the
manner shown in Figure \ref{wavy} for the entire
duration of the integration, 3 Myr.
The libration centers are $\langle \phi_{5:2} \rangle = 180\degr$,
the amplitudes are $\Delta \phi_{5:2} \equiv \max \phi_{5:2} - \langle
\phi_{5:2} \rangle \approx 90\degr$--$140\degr$,
and the libration periods are $T_l \approx 2 \times 10^4\yr$.
The libration period increases with decreasing libration amplitude,
unlike the case for the conventional pendulum model for a resonance.

\placefigure{fig3}

For each object, we further explore error space by integrating 8 additional
sets of initial conditions that lie on the 2$\sigma$,
3$\sigma$, 4$\sigma$, and 5$\sigma$ confidence surfaces and that
are characterized by semi-major
axes that deviate most from the best-fit semi-major axis
in positive and negative senses.
Objects (38084) 1999HB$_{12}$ and 2001KC$_{77}$ remain in the 5:2 $e^3$
resonance in all cases for 3 Myr. We conclude that our identifications
of (38084) 1999HB$_{12}$ and 2001KC$_{77}$
as current 5:2 librators are particularly secure.
Object 1998WA$_{31}$ fails to librate in the 5:2 resonance
when its initial semi-major axis is less than the nominal
value by 2$\sigma$ or more, i.e., when $a$ is less than
the nominal value by more than 0.13 AU. However, other
sets of initial conditions for
which $a$ is greater than the nominal value
yield libration even at the 5$\sigma$ level for 1998WA$_{31}$.
Our identification of 1998WA$_{31}$ as a current member
of the resonance is therefore less firm than for the others,
but not alarmingly so.

What is the long-term evolution of these objects? We have
integrated orbit solutions 1, 2, and 3 for all three objects
forward by 1 Gyr. For (38084) 1999HB$_{12}$, all 3 orbit solutions
yield libration in the 5:2 resonance for the full
duration of the integration. The same is true for orbit
solution 2 of 2001KC$_{77}$. For the aforementioned
4 trajectories, the libration amplitudes range from
$90\degr$ to $100\degr$. By contrast, solution 1 of 2001KC$_{77}$
eventually yields circulation, while solution 3 leads to a close
encounter with Neptune 0.512 Gyr into the simulation.
For 1998WA$_{31}$, all three solutions eventually end with
a close encounter with Neptune, with solution 2 lasting the
longest (0.882 Gyr). We conclude that among our three 5:2 resonant
members, (38084) 1999HB$_{12}$ and possibly 2001KC$_{77}$
are likely to be long-term and therefore primordial residents of
the 5:2 MMR. Note further that the accuracy of our orbital solution
is highest for (38084) 1999HB$_{12}$ and lowest
for 1998WA$_{31}$; thus, it seems possible that
with more astrometric measurements, all 3 objects will be found to stably
occupy the 5:2 MMR over Gyr-long timescales.

Orbital elements for our three 5:2 resonant KBOs are provided
in Table 1.

\subsection{Observed Member of the 1:1 MMR}
\begin{figure}
\epsscale{0.85}
\vspace{-0.2in}
\plotone{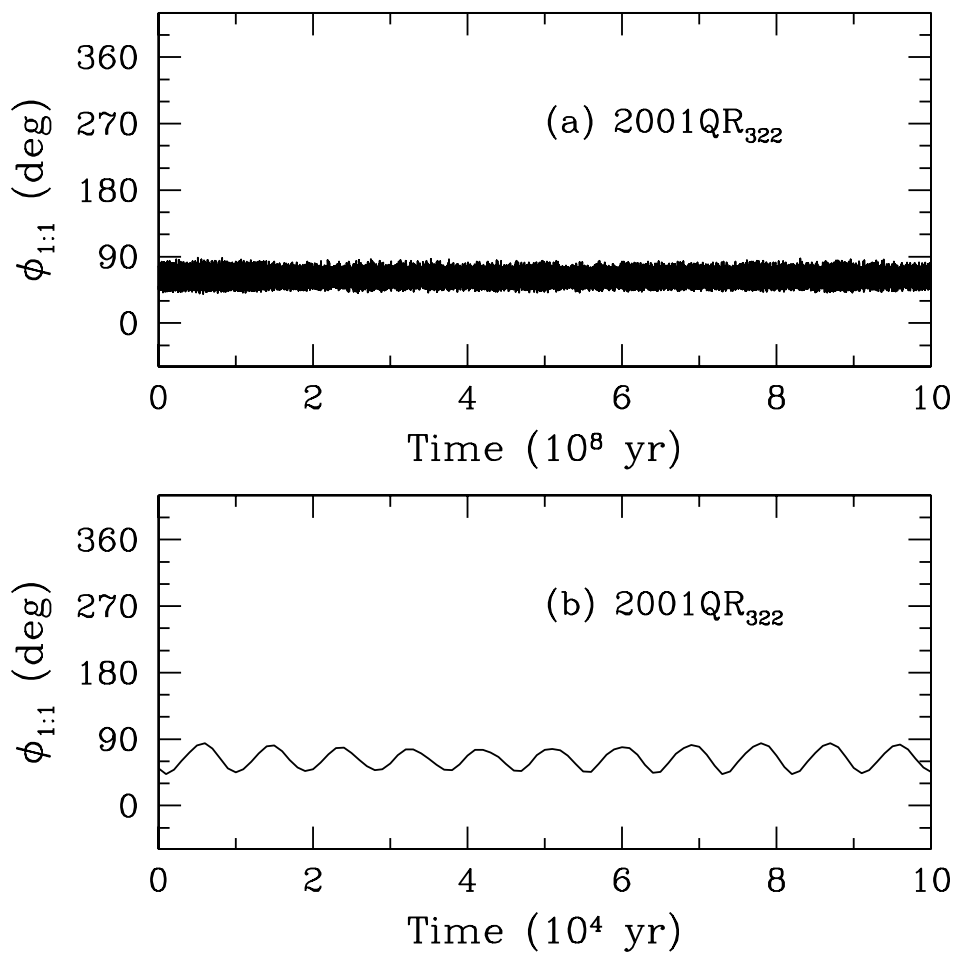}
\caption{Evolution of the resonant argument, $\phi_{1:1} = \lambda -\lambda_N$,
for our Neptunian Trojan, based on best-fit orbit solution 1.
The object remains bound to the 1:1 resonance for
at least 1 Gyr and betrays no sign of instability.
Top and bottom panels display the same evolution
with different time resolutions.
Orbit solutions that deviate from the best-fit
solution by as much as 5$\sigma$ also yield libration for at least 3 Myr (data
not shown).}
\label{wavytrojan}
\end{figure}

The evolution of the resonant argument, $\phi_{1:1}$, for object
2001QR$_{322}$ is displayed in Figure \ref{wavytrojan}. Only
the integration of the best-fit solution is shown;
orbit solutions 2 and 3 yield nearly identical results.
All three sets of initial conditions yield tadpole-type libration
about Neptune's L4 point for at least 1 Gyr.
The libration center is $\langle \phi_{1:1} \rangle \approx 64\fdg 5$,
the libration amplitude is $\Delta \phi_{1:1} \approx 24\degr$,
and the libration period is $T_l \approx 10^4\yr$.

\placefigure{fig4}

We further explore error space by
integrating 8 additional solutions that deviate from the nominal best-fit
solution by 2$\sigma$, 3$\sigma$, 4$\sigma$, and 5$\sigma$,
each for 3 Myr. In all cases tested, object 2001QR$_{322}$ librates
in the 1:1 MMR. We regard our identification of 2001QR$_{322}$
as a Neptunian Trojan as extremely secure.

\begin{figure}
\epsscale{0.85}
\vspace{-0.6in}
\plotone{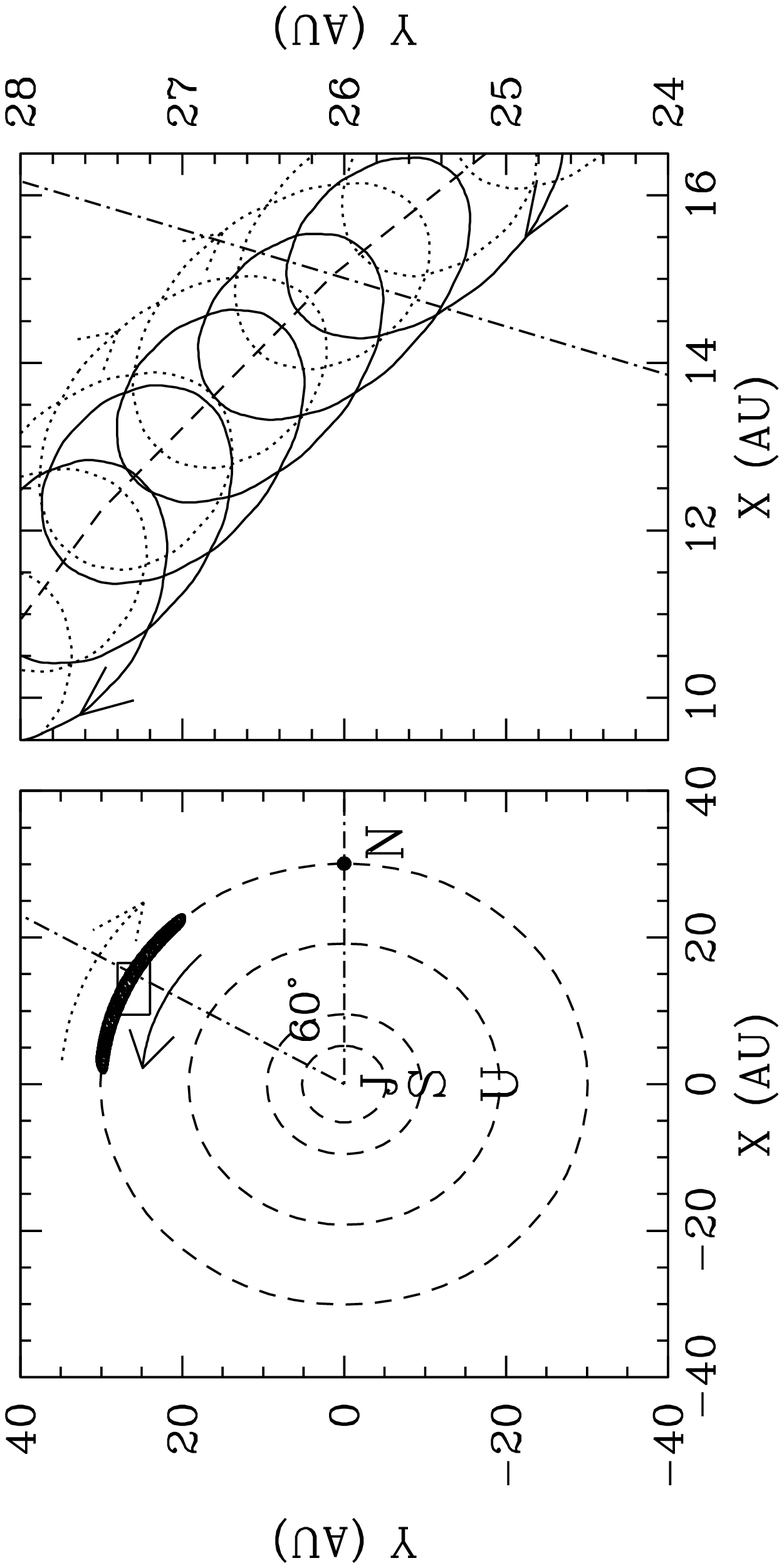}
\caption{Trajectory of 2001QR$_{322}$, our Neptunian Trojan, in a
quasi-Neptune-centric frame. The left-hand panel displays a bird's-eye
view of the outer solar
system, with the giant planet orbits
shown schematically. The dark tube of points lying
on Neptune's orbit marks the computed path of the Trojan.
The length of the vector from the origin to each point on the tube
gives the instantaneous heliocentric distance of the object;
the angle between this vector and the
abscissa gives the instantaneous
angle between the Sun-Trojan and Sun-Neptune vectors.
The Trojan librates along Neptune's orbit as indicated
by the solid and dotted curved arrows. Each libration takes about 10$^4$ yr
to complete. The small inset rectangle is magnified in the right-hand
panel to show the fast epicyclic motion. Each fast epicycle
takes about 1 orbital period of Neptune, or about 200 yr, to complete.
}
\label{trojanxyz}
\end{figure}

The trajectory of the Trojan in the Neptune-centric frame
is showcased in Figure \ref{trojanxyz}. A tadpole-like path
whose center is shifted forward in longitude
from Neptune's L4 point is evident; the longitude shift of
$\sim$5$\degr$ is expected for
finite amplitude librators (see, e.g., Murray \& Dermott 1999, their
Figure 3.11). The minimum distance of approach to Neptune over
3 Myr is approximately 20 AU.

\placefigure{fig5}

Orbital elements for our Trojan are listed in Table 1.
Note that the orbital elements of 2001QR$_{322}$
lie consistently within the region of 4-Gyr-long stability
mapped by Nesvorny \& Dones (2002); see their Figure 9c.
Our object is a member of the low-inclination population
of stable Neptunian Trojans; Nesvorny \& Dones (2002)
find surprisingly that Neptune Trojans having orbital inclinations
as high as $25\degr$ are also stable. Note further
that the libration amplitude of 2001QR$_{322}$
($24\degr$) also lies consistently below the stability
threshold of 60--$70\degr$ established by Nesvorny \& Dones (2002).

How many Neptunian Trojans might there be in all?
Nesvorny \& Dones (2002) provide three models of the sky
density of Neptune Trojans that differ in the assumed
distribution of orbital elements. We have combined
their models (see their Figures 11, 13, and 14) with
the distribution of our DES search fields to estimate
that $\sim$20, $\sim$60, and $\sim$40 Neptune Trojans
having diameters and albedos
comparable to those of 2001QR$_{322}$
exist in all, based on their models I, II, and III, respectively.
The above numbers already include Trojans
librating about Neptune's L5 point, which we assume
to have the same population as L4 librators.
While it is impossible to differentiate between the models
based only on the discovery of a single object, it is
heartening to see that all 3 models give the same
order-of-magnitude estimate for the total number of Neptune Trojans
resembling 2001QR$_{322}$.

\section{Theoretical Implications of Resonance Occupation}
\label{models}
Here we briefly explore theoretical implications of the observed
occupation of the 5:2 and 1:1 Neptunian MMRs. We aim, in particular,
to test the hypothesis that Neptune migrated outwards by several AU
during the solar system's past and, in so doing, sculpted
the pattern of resonance occupation in the Kuiper belt (Fernandez \& Ip 1984;
Malhotra 1995; CJ). Sections \ref{mighyp} and \ref{nomig}
focus on the 5:2 MMR, while section \ref{tt} is devoted to the 1:1 MMR.

\subsection{Neptune's Migration and the 5:2 MMR}
\label{mighyp}

Could KBOs in the 5:2 resonance have been trapped
into that MMR as it swept across the Kuiper belt?
We consider two scenarios, one in which
Neptune migrates into a sea of initially dynamically
cold test particles, and another in which
the planet migrates into a sea of initially dynamically
hot particles.

\subsubsection{Cold Initial Conditions}
\label{cic}

For objects on initially low-eccentricity orbits,
the probability of capture into the 5:2, third-order
resonance is prohibitively small compared to the probability
of capture into low-order resonances such as the
2:1 and 3:2. Neither the
simulations performed by Malhotra et al.~(2000),
nor those by CJ report any
object caught into the 5:2 resonance among the
$\sim$100 test particles over which that resonance
swept. We have executed another migration simulation,
following those of CJ, that is
tailored to gauge the capture efficiency of the
5:2 resonance for objects on initially nearly
circular, low-inclination orbits. The simulation parameters are identical
to those in CJ's model I, except
that the initial semi-major axes of the 400 test
particles range from 43.55 AU (= 1 AU greater than
the initial location of the 5:2 resonance)
to 54.44 AU (= 1 AU less than the final location
of the 5:2 resonance). Thus, all such objects
are potentially swept into the migrating 5:2 resonance.
Their initial eccentricities and inclinations are
randomly and uniformly distributed between 0.00 and 0.05,
and between 0.00 and 0.025 rad, respectively. Arguments of periastron
($\omega$), longitudes of ascending nodes ($\Omega$),
and mean anomalies ($M$) are uniformly and randomly sampled
between 0 and $2\pi$. The semi-major
axis of each giant planet evolves with time, $t$, according to

\begin{equation}
a(t) = a_f - (a_f - a_i) \exp{(-t/\tau)} \, ,
\end{equation}

\noindent where we fix the migration timescale, $\tau$,
to be $10^7\yr$. We adopt values for
the initial and final semi-major axes,
$(a_i,a_f)$, for each of the planets as follows (in AUs): Jupiter
$(5.40,5.20)$, Saturn $(8.78,9.58)$, Uranus $(16.2,19.2)$,
and Neptune $(23.1,30.1)$. We employ the symplectic
integrator, SyMBA (Duncan, Levison, \& Lee 1998),
as kindly supplied to us by E.~Thommes.
We adopt a timestep of 0.6 yr.
For more details, the reader is referred to CJ.

Note that our simulations prescribe the migration to be smooth.
If the planetesimals that scattered off Neptune and drove
its migration were sufficiently massive, our idealization would
be invalid. We estimate that our approximation is valid
if most of the mass of the planetesimal disk were contained
in bodies having radii less than $\sim$40 km.
The derivation of our crude estimate is contained in
the Appendix. The actual sizes of ancient planetesimals
scattering off Neptune are, of course, unknown, though
Kenyon (2002) calculates in his accretion simulations
that $\sim$90\% of the solid mass at heliocentric distances
of 40--50 AU in the primordial solar system may be
contained in 0.1--10 km-sized objects.

\begin{figure}
\epsscale{0.85}
\vspace{-0.5in}
\plotone{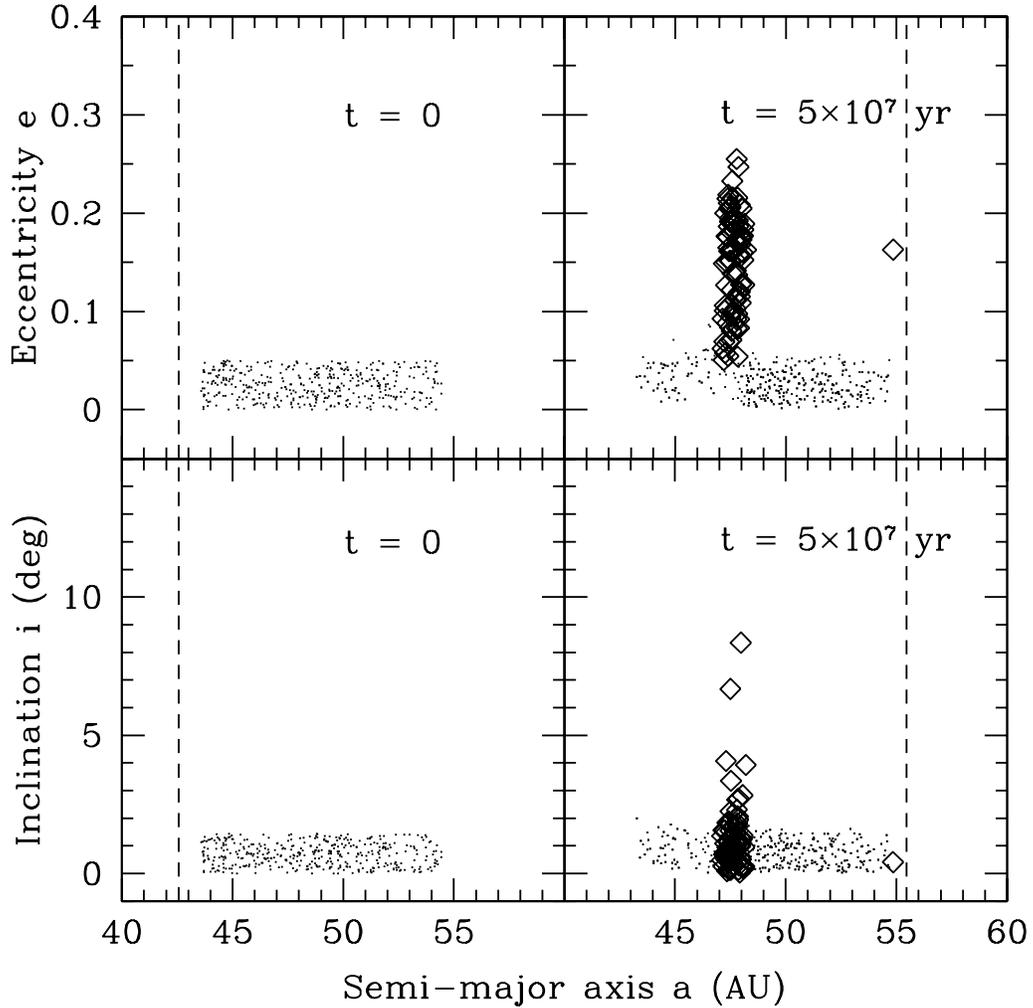}
\caption{
Results of a migration simulation designed to gauge
the capture efficiency of the 5:2 resonance. Left-hand panels
display cold initial conditions of 400 test particles prior
to sweeping by the 5:2 resonance, whose nominal
location is indicated by the dashed line. Right-hand panels
display the aftermath of resonance sweeping, where open
diamonds denote resonantly librating particles. Only 1 particle
is caught by the 5:2 MMR; its eccentricity is pumped to
0.16 and its inclination is relatively unaltered.
By contrast, 90 particles are swept into the 2:1 resonance
at $a \approx 47.8 \AU$. The ratio of 1-to-90 is difficult to reconcile
with the observed 3-to-1 ratio showcased in Figure \ref{ae};
accounting for the bias introduced by the fact that the 5:2 MMR is more distant
than the 2:1 would only worsen the disagreement. Moreover,
the predicted final $e$ and $i$ of our simulated 5:2 resonant object
are much lower than observed values.
}
\label{frame52}
\end{figure}

Figure \ref{frame52} demonstrates
that capture into the 5:2 resonance, even when
Neptune takes as long as a few $\times$ $10^7 \yr$
to migrate outwards by several AU, is
improbable; only 1 out of 400 objects librates
in the 5:2 resonance at the end of the simulation.
By contrast, the 2:1 resonance boasts 90 captured
objects. The predicted population ratio between the 5:2
and 2:1 resonances is not easily reconciled with
the observations as depicted in Figure \ref{ae}.
Accounting for the observational bias in favor
of finding 2:1 members over 5:2 members
due to the fact that the 5:2 resonance is more distant than the 2:1
would only accentuate the disagreement.
When both the effects of greater distance and differential
longitudinal clustering of resonant KBOs are accounted for,
we estimate that bias correction factors of $\sim$3 in favor
of finding 2:1 members result (see CJ for a discussion
of how these bias corrections are estimated).
Even if the difference between the predicted
1-to-90 ratio vs. the observed 3-to-1 ratio
were to be attributed
to extremely strong and positive radial gradients in the primordial surface
density of planetesimals (an unnatural prospect in itself),
resonant excitation by the 5:2 MMR of initially cold orbits results
in eccentricities and inclinations that are generally much too low
compared with the observations.

\placefigure{fig6}

Finally, we note that when non-DES and DES
datasets are combined, the number of securely identified 2:1 resonant KBOs
increases to 6. Blithely using this number, which is affected by
observational biases from non-DES surveys that we have not quantified, still
yields a ratio (3-to-6) that is hard to reconcile
with the predicted ratio (1-to-90).
The 6 2:1 resonant objects are
(20161) 1996TR$_{66}$,
(26308) 1998SM$_{165}$, 1997SZ$_{10}$, 1999RB$_{216}$, 2000JG$_{81}$, and
2000QL$_{251}$. Membership of 1997SZ$_{10}$ in the 2:1 resonance,
and suspicion of membership of (20161) 1996TR$_{66}$,
were previously reported by Nesvorny \& Roig (2001).

\subsubsection{Hot Initial Conditions}
\label{hic}

\begin{figure}
\epsscale{0.85}
\vspace{-0.2in}
\plotone{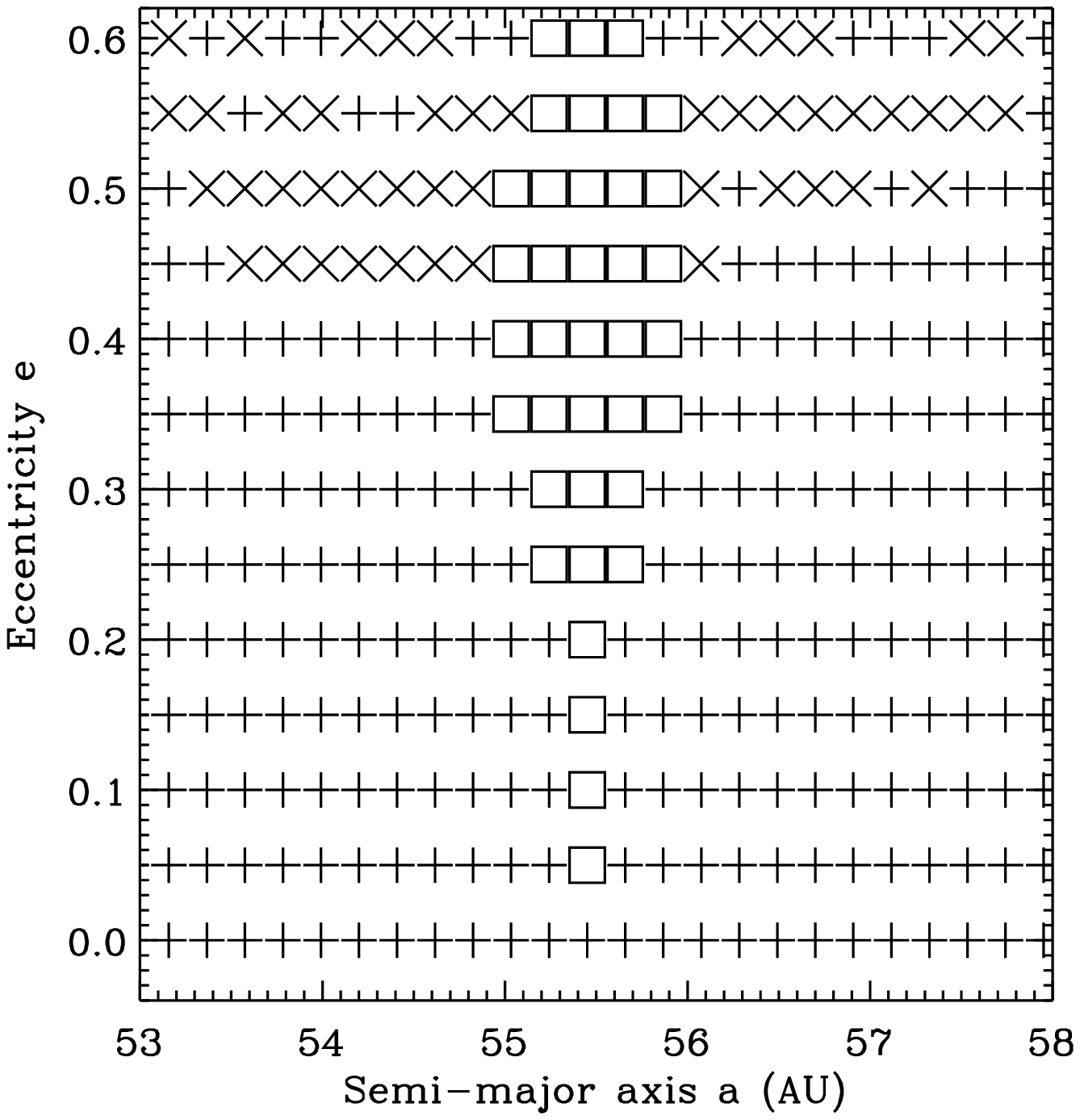}
\caption{Estimated width of the 5:2 MMR, derived by numerical integration
of the circular, planar, restricted 3-body model for the
Sun/Neptune/KBO system. At each point on the above grid
in $a$-$e$ space, a test particle's trajectory
is integrated in the gravitational fields of the Sun and Neptune
for $3\times 10^5\yr$ using a timestep of 0.6 yr.
Test particles share the same initial $i=0$,
$\Omega = 0$, $\omega = \pi/2$, and $M = 0$. Initial elements of
Neptune are given by
$a_N = 30.1 \AU$, $e_N = i_N = \Omega_N = \omega_N = M_N = 0$.
Thus, each particle's initial $\phi_{5:2} = \pi$. Plus signs (``$+$'')
denote particles for which $\phi_{5:2}$ circulates; crosses (``$\times$'')
denote particles that encounter the Hill sphere of Neptune;
and open squares denote particles that librate in the 5:2 MMR.
The resonant width is greatest, $\Delta a \approx 0.8 \AU$,
at $e\gtrsim 0.2$. Our 3-body model
is used only to generate this figure and for no other figure
in this paper.}
\label{width}
\end{figure}

Figure \ref{width} displays the width of the 5:2
resonance in $a$-$e$ space. Since the resonance
widens considerably at $e \gtrsim 0.2$, it
is worth considering whether the capture efficiency
increases with increasing initial eccentricity.
Hot initial conditions prior to Neptune's
planetesimal-induced migration would be expected
under the scenario of Thommes, Duncan, \& Levison (1999, 2002).
In their scenario, proto-Neptune and proto-Uranus are
scattered by the other, nascent
giant planets to heliocentric distances beyond 30 AU and heat
the primordial Kuiper belt by gravitational scattering.
That the Kuiper belt has been disturbed by more
than the (hypothesized) slow sweeping of Neptune's MMRs is evidenced
by KBOs' large orbital inclinations (Brown 2001; CJ).

\placefigure{fig8}

\begin{figure}
\epsscale{0.85}
\vspace{-0.5in}
\plotone{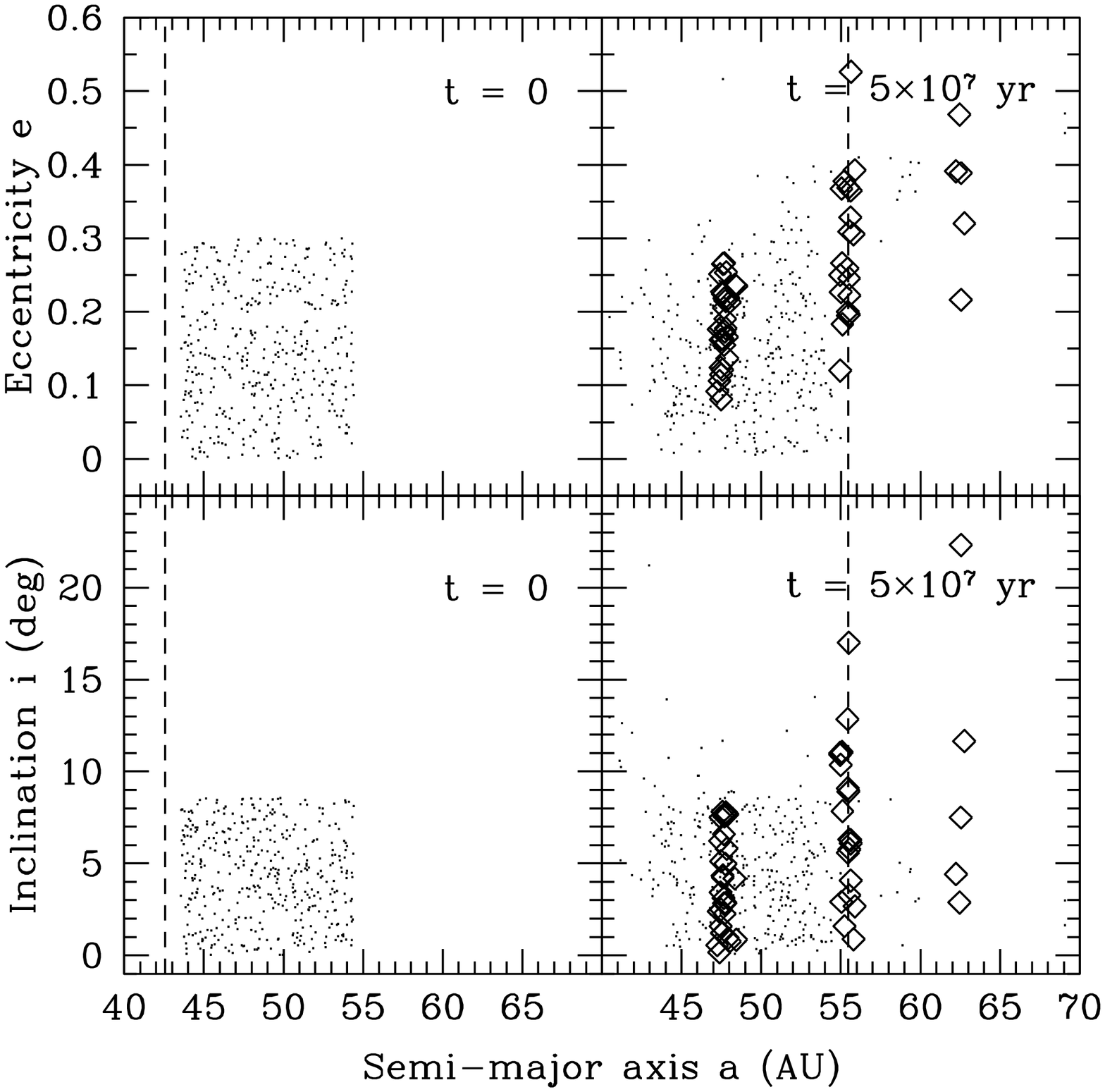}
\caption{Results of a migration simulation designed to gauge
the capture efficiency of the 5:2 resonance
under hot initial conditions. Left-hand panels
display hot initial conditions of 400 test particles prior
to sweeping by the 5:2 resonance, whose nominal
location is indicated by the dashed line. Right-hand panels
display the aftermath of resonance sweeping. Open diamonds
denote librating particles in the 2:1, 5:2, and 3:1 MMRs.
Twenty particles
are adiabatically swept into the 5:2 MMR and have their eccentricities
and inclinations excited above their initial values.
Twenty-nine particles are swept into the 2:1 resonance
at $a \approx 47.8 \AU$. The libration amplitudes of the
simulated 5:2 resonant KBOs range from $16\degr$ to $145\degr$,
with 6 particles having amplitudes in the range between
$90\degr$ and $140\degr$ (data not shown).
The relative efficiencies of capture into the 2:1 and 5:2 MMRs,
the predicted libration characteristics of 5:2 resonant particles,
and the large final eccentricities and inclinations of resonant
particles
can all be reconciled with the observations, in contrast to the case
using only cold initial conditions.
}
\label{frame52x}
\end{figure}

We repeat the migration simulation of \S\ref{cic}
but with initial eccentricities and inclinations
of test particles uniformly and randomly distributed between $0$ and $0.3$,
and between $0$ and $0.15$ rad, respectively. The result
is summarized in Figure \ref{frame52x}. Of 400
particles potentially caught by the sweeping 5:2 MMR,
20 are captured and have their eccentricities
amplified to final values of 0.2--0.5. We have verified that these 20
objects represent adiabatic capture events
and not violent scatterings; their semi-major axes
increase smoothly over the duration of the simulation
from values of as low as 45 AU to the final resonant value
of 55.4 AU.

\placefigure{fig9}

In addition, 5 objects are adiabatically swept into the 3:1 resonance
whose final location lies at $a \approx 62$ AU.

Are the predicted libration amplitudes consistent with those
observed? The answer is yes; libration amplitudes of our 3 observed
KBOs range from $90\degr$ to $140\degr$, while
those of our 20 simulated 5:2 resonant particles range from $16\degr$
to $145\degr$, with 6 particles having amplitudes
in the observed range.

While the capture efficiencies of high-order MMRs such as the
5:2 and 3:1 resonances magnify with increasing initial
eccentricity, those of low-order MMRs such as the 3:2 and 2:1
resonances decrease. In the simulation just described,
29 objects are caught in the 2:1 resonance---a factor of 3
decline over the case with cold initial conditions.
We have verified that these 29 objects are adiabatically captured
by the sweeping 2:1 resonance and are not scattered
into it by close encounters with one of the giant planets.
These captured particles
originated on low eccentricity orbits, $e \lesssim 0.1$.

Taken at face value, one problem with the simulation
depicted in Figure \ref{frame52x} is that it predicts
a large proportion of non-resonant particles having semi-major
axes between 50 AU and 55 AU that, to date, are not observed.
The problem of the ``Kuiper Cliff''---a sudden decrease
in the surface density of planetesimals outside 50 AU---has been discussed
extensively in the literature (see, e.g., Jewitt, Luu, \& Trujillo~1998;
Gladman et al.~1998; Chiang \& Brown~1999; Allen, Bernstein, \& Malhotra~2001;
Trujillo \& Brown~2001). We regard the statistical significance
of the observed edge of the belt as still marginal at best
(see Allen et al.~2001). But even apart from possible observational
selection biases, there are a number of factors that would help
to improve the agreement between Figure \ref{frame52x}
and observation. First, a fraction
of the simulated non-resonant objects between
$a = 50\AU$ and 55 AU have large eccentricities and are not phase-protected
from Neptune, so that they are unlikely to survive in their current orbits
for the age of the solar system. Second, if the ``Kuiper Cliff''
is real, we may impose an edge to our distribution
at $a = 50\AU$ prior to resonance sweeping
that would obviously reduce the number of objects
in this region after resonance sweeping. The number of objects
caught in the 5:2 MMR would be reduced by $\sim$50\% compared
to that shown in Figure \ref{frame52x}.
The resultant population ratio between the 5:2 and 2:1 MMRs of $\sim$10-to-29
would still be reconciliable with the observations
if we include the 5 other non-DES 2:1 objects
listed in \S\ref{cic}.

In summary, slow resonance sweeping over a primordial
Kuiper belt that comprises both pre-heated orbits
having $i,e \gtrsim 0.2$ and cold orbits having $e\lesssim 0.1$
is able to populate the 2:1 and 5:2 MMRs
with efficiencies that do not seem irreconciliable with the
observations. The models can be tuned to
match the observations by adjusting the initial eccentricity,
inclination, and semi-major axis distributions of belt particles
prior to the migration phase. We have not undertaken such
tuning here; our main conclusion is that capture into the
5:2 MMR is made substantially more efficient, and generates
5:2 resonant orbits similar to those observed, by pre-heating the belt
prior to resonance sweeping.
Of course, our finding does not address the question
of what was responsible for this pre-heating.

\subsection{Direct Scattering into the 5:2 MMR}
\label{nomig}

Is it possible that KBOs in the 5:2 MMR may not have
been captured via resonance sweeping, but were rather
gravitationally scattered into that resonance by close
encounters with one or more massive objects?
In $a$-$e$-$i$ space, the proximity of our 5:2 resonant
KBOs to orbits traditionally described as ``scattered''
suggests direct scattering by Neptune as a population mechanism.
To test this hypothesis, we integrate the trajectories
of 400 test particles on initially low-eccentricity, low-inclination
orbits in the vicinity of Neptune. The test particles'
semi-major axes, eccentricities, and inclinations
range between 31.7 AU and 35.7 AU (2--7 Neptunian
Hill radii from Neptune's semi-major axis), 0.00 and 0.02,
and 0.00 and 0.01 rad, respectively. The other orbital angles
are uniformly and randomly distributed between 0 and $2\pi$.
The duration of the integration
is $5 \times 10^7\yr$. No migration is imposed
on any of the planets, whose initial positions and velocities
are taken from Cohen, Hubbard, \& Oesterwinter (1973).
We again employ the symplectic integrator, SyMBA;
this integrator handles close encounters
with better accuracy than does swift\_rmvs3.
We adopt a timestep in the absence of close
encounters of 0.6 yr.

\begin{figure}
\epsscale{0.85}
\vspace{-0.5in}
\plotone{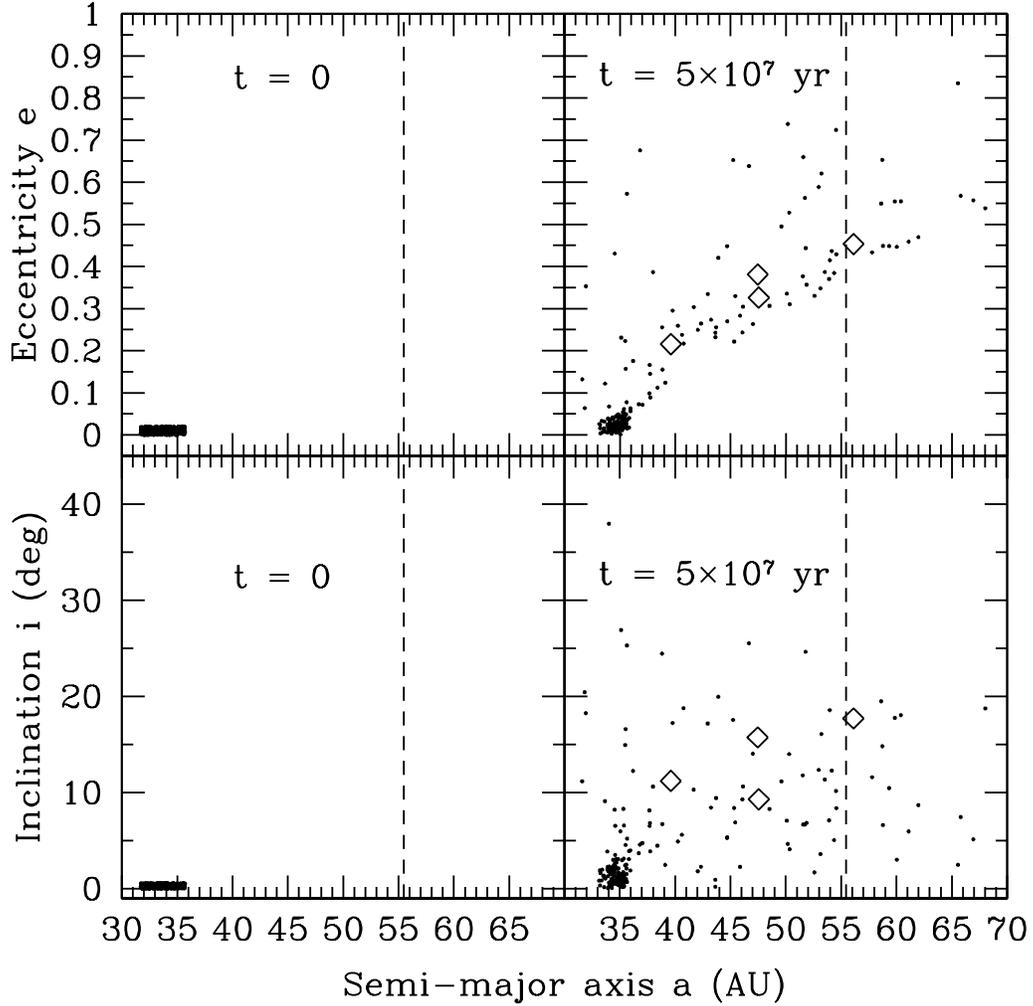}
\caption{Results of a direct scattering simulation
in which no migration is imposed on any of the planets. Left-hand
panels display cold initial conditions for 400 test particles
situated between 2 and 7 Neptunian Hill radii from Neptune.
Right-hand panels display conditions after 50 Myr.
One, two, and one particles, represented by open diamonds,
are scattered directly into the
5:2, 2:1, and 3:2 MMRs, respectively. The eccentricities,
inclinations, and relative numbers of these particles
appear consistent with the observations. However, the predicted
and observed libration
characteristics disagree. The four simulated resonant particles
have libration amplitudes between $160\degr$ and $175\degr$ (data
not shown in this figure), too large compared to the moderate
amplitudes observed.}
\label{scat}
\end{figure}

Figure \ref{scat} summarizes the results of this simulation
of direct scattering into MMRs. Of 400 test particles,
0, 1, 2, and 1 particles are scattered into the 3:1, 5:2,
2:1, and 3:2 MMRs. These resonant KBOs have
substantial eccentricities, between 0.19 and 0.45.
Thus, the existence of resonant
KBOs having high eccentricities, taken at face value, does not necessarily
imply capture and adiabatic excitation
by migratory resonances.
The relative capture efficiencies between the 3:1, 5:2, 2:1,
and 3:2 MMRs in our direct scattering simulation
do not appear irreconciliable with the observations,
given the small number statistics (both
observationally and theoretically), observational biases
(CJ), and uncertainties regarding actual initial conditions.

Direct scattering by Neptune, however, predicts
libration amplitudes that are generally
larger than those observed. For our (4) simulated
resonant particles inhabiting the 5:2, 2:1, and 3:2 MMRs,
libration amplitudes all exceed $160\degr$. This is to be compared, for
example, with the libration amplitudes exhibited by our
three observed 5:2 resonant KBOs, which range
between $90\degr$ and $140\degr$ (see Figure \ref{wavy}).
While we cannot rule out the possibility that 1998WA$_{31}$
represents such a directly scattered, dynamically young object based
on its unstable behavior on timescales of Myrs to Gyrs (see \S\ref{fivetwo}),
the small libration amplitudes of (38084) 1999HB$_{12}$ and 2001KC$_{77}$
do not match those predicted by the scattering simulation.
(And as noted in \S\ref{fivetwo}, it remains possible that future
improvements in the accuracy of the trajectory of 1998WA$_{31}$
may cause it to join its more stable brethren.)

The problem of excessive libration
amplitudes was reported in a similar context by Levison \& Stern (1995),
who investigated ways to excite Pluto's orbit to its present
high eccentricity and inclination using only gravitational
interactions with the giant planets in their current orbits.
Possible resolutions to this difficulty include invoking physical
collisions with and/or gravitational scatterings off primordial
KBOs (Levison \& Stern 1995).
The disagreement between
predicted and observed libration
characteristics seems particularly severe
for the 2:1 resonant objects. In our direct scattering simulation,
the two particles scattered into the 2:1 MMR librate with large
amplitude about $\langle \phi_{2:1} \rangle = 180\degr$. This conflicts
with the observed
small libration amplitudes about $\langle \phi_{2:1} \rangle \approx
75\degr$:
the single confirmed Twotino in our survey librates with small
libration amplitude ($50\degr$) about
$\langle \phi_{2:1} \rangle \approx 88\degr$, while
four secure non-DES Twotinos are characterized by
$\langle \phi_{2:1} \rangle \approx 70\degr, 67\degr, 74\degr,$
and $83\degr$ (see CJ).
Only one secure non-DES Twotino (2000JG$_{81}$)
resembles a simulated particle, librating about $\langle \phi_{2:1} \rangle =
180\degr$
with an amplitude of $160\degr$.
Note that we interpret the observed asymmetrically librating Twotinos
to be primordial residents, since they resemble the stable
particles simulated by Nesvorny \& Roig (2001; see their section 3.4).

Is it possible that our simulation
contains too few particles to fully explore phase space,
and that we have been unlucky in the outcome of libration
profiles? We do not believe so. Objects barely bound to
MMRs are to be expected from the direct scattering hypothesis,
because for Neptune to heat the orbit of a test
particle significantly, the distance of closest
approach must be small, within several Hill radii
of Neptune. During the (brief) close encounter, a particle's velocity
is radically altered; but because the particle's position during
the encounter is relatively unchanged, subsequent close approaches
between particle and planet will occur at similarly small distances. Large
libration amplitude is synonymous with small distance of
closest approach; thus, close encounters with Neptune
alone are expected to yield only tenuously bound resonant
particles whose resonant locks are easily broken.

\placefigure{fig10}

\subsection{Neptune's Migration and the 1:1 MMR}
\label{tt}

Gomes (1998) explores, via numerical simulations similar to those
presented here, the ability of a migratory Neptune to
{\it retain} (as opposed to capture) a Trojan population.
A variety of migration histories for the giant planets
are tested; between 20\% and 82\% of his hypothetical Neptunian Trojans
remain bound to the 1:1 MMR throughout the migration phase. He concludes
that unless the migration history of the giant planets was such
as to engender divergent resonance crossings and excitation
of planetary eccentricities to values of $\sim$0.1---a history
that would, {\it prima facie}, conflict with the small orbital
eccentricities currently exhibited by Uranus and Neptune---a Neptunian
Trojan population might be expected to exist today.
This finding is strongly reinforced by Nesvorny \& Dones (2002),
who find for their hypothesized Neptune Trojans that
about 50\% of them survive for 4 Gyr in a post-migration solar system.

\begin{figure}
\epsscale{0.85}
\vspace{-1.5in}
\plotone{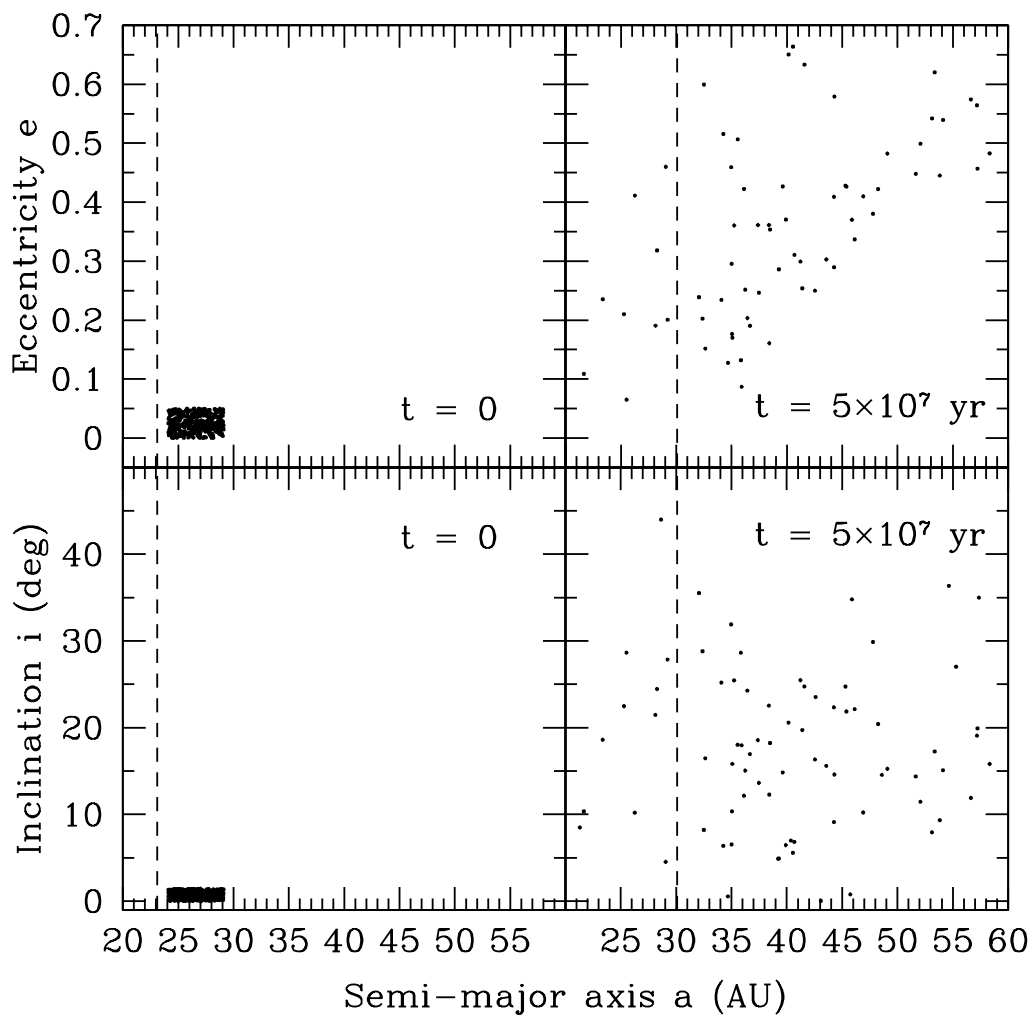}
\caption{
Results of a migration simulation designed to gauge
the capture efficiency of the 1:1 resonance. Left-hand panels
display cold initial conditions of 400 test particles prior
to sweeping by the 1:1 resonance, whose nominal
location is indicated by the dashed line. Right-hand panels
display the aftermath of resonance sweeping. No particle
is captured into a Trojan-type orbit. Particles
instead suffer close encounters with Neptune and
are scattered onto highly eccentric and inclined orbits.
}
\label{frame11}
\end{figure}

Distinct from the retainment efficiency of the 1:1 MMR
is the efficiency with which a migrating Neptune captures objects into the 1:1
MMR; this capture efficiency has not been reported in the literature.
To remedy this deficiency, we execute a migration simulation
similar to the one described in \S\ref{cic},
except that the initial semi-major axes of the 400 test
particles are distributed between 24.1 AU (= 1 AU greater than the
initial location of the 1:1 MMR) and 29.1 AU (= 1 AU less than the
final location of the 1:1 MMR). Only cold initial
conditions (initial $0 \leq e \leq 0.05$, $0 \leq i \leq 0.025$)
are employed; hot initial conditions would conflict with the
observed low $e$ and low $i$ of 2001QR$_{322}$. This is because, as described
in greater detail below, migration is not expected to significantly
alter the eccentricities and inclinations of Trojans. Results
of the migration simulation are displayed in Figure \ref{frame11}.
Of 400 particles potentially swept into the 1:1 MMR, no particle is
captured. The overwhelming fate of the particles is to be scattered
to larger semi-major axes, eccentricities, and inclinations.
The low capture efficiency of $\lesssim 0.0025$ suggests
that Neptunian Trojans do not owe their existence to Neptune's
migration. Our conclusion is subject to the caveat that
we have not modelled a possible stochastic component
to Neptune's migration; see \S\ref{cic} and the Appendix.

Capture scenarios that invoke
frictional drag from solar nebular gas, and damping of libration
amplitudes via gaseous envelope accretion by the host planet, are
characterized by healthy capture efficiencies for Jupiter's Trojans
(Marzari \& Scholl 1998; Peale 1993), but are expected to be
less efficient for Neptune's Trojans. The decrease in efficiency
arises because Neptune's hydrogen/helium component amounts to only
$\sim$5--20\% of Neptune's total mass, while Jupiter's hydrogen/helium
component comprises $\sim$90\% of that planet's mass (Lissauer 1995).
We also note that frictional drag was likely to have been
ineffective if 2001QR$_{322}$ possessed its current diameter (estimated
130--230 km, assuming a 12--4\% albedo) at the time of capture;
2001QR$_{322}$ must have grown from the collisional
agglomeration of smaller bodies that did feel gas drag
(Peale 1993).\footnote{We
note in passing that Neptune's irregular satellite, Triton, might be thought of
as inhabiting a
1:1 resonance with Neptune and might have been captured
from an initially heliocentric orbit by colliding with an ancient
regular satellite of Neptune (Goldreich et al.~1989).}

\placefigure{fig11}

We now justify our earlier assertion that hot initial conditions
prior to resonance capture into the 1:1
resonance are inappropriate for 2001QR$_{322}$.
Unlike the case of exterior resonances, outward migration
causes the eccentricities and inclinations of Trojans to decline.
This behavior can be seen from the adiabatic invariant, $C_{pq}$,
associated
with a MMR for which the ratio of mean orbital
periods is $p$:$(p+q)$ ($p$ and $q$ are integers
and $q < 0$ for exterior resonances):

\begin{eqnarray}
C_{pq} & = & \sqrt{M_{\odot} a} \, [(p+q) - p\sqrt{1-e^2} \cos i] \\
 & \approx & \frac{\sqrt{M_{\odot} a}}{2} \, (e^2 + i^2) \, .
\end{eqnarray}

\noindent (see, e.g., Yu \& Tremaine 2001).
For the last equality, we have taken
$p = 1$, $q = 0$, and $e,i \ll 1$. Then
as $a$ increases, $e^2 + i^2$ must decrease. Fleming \& Hamilton (2000)
find in numerical simulations that, indeed, both eccentricity and
inclination decrease as the semi-major axis increases, in quantitative
agreement with the adiabatic invariant. The effect
is extremely weak. If we take $i^2$ to be always comparable to $e^2$
(as we can for the current orbit of 2001QR$_{322}$), then $(e_f/e_i) \approx
(i_f/i_i) \approx (a_i/a_f)^{1/4}$.
Migration scenarios adopt, for Neptune, $0.7 \lesssim a_i/a_f \leq 1$;
then the eccentricities and inclinations
of its Trojans decline by at most 10\% due
to outward migration.

Perhaps the most natural time for Neptune to accrue its Trojans is
during its assembly from planetesimals. For mass accretion timescales
that are long compared to the Trojan libration period, a 30-fold increase
in the mass of the host planet reduces the libration amplitudes
of Trojans by factors of 2--3 (Fleming \& Hamilton 2000).

\section{Summary and Discussion}
\label{sum}

We have presented, as part of our ongoing Deep Ecliptic Survey,
the most complete picture of mean-motion resonance occupation
in the Kuiper belt to date (Figure \ref{ae}). We have discovered
members of the 5:2, 2:1, 7:4, 3:2, 4:3, and 1:1 resonances.
These KBOs represent secure identifications in the sense
that (1) their 1$\sigma$ fractional uncertainties in semi-major
axis lie between 3\% and 0.003\%, and
(2) numerical integrations of orbit solutions distributed
over the $1\sigma$ confidence surface of possible fitted
orbits consistently yield resonant arguments that librate for at least 3 Myr.
In the special cases of the 1:1 and 5:2 resonances, we have checked by
explicit numerical orbit integrations
that orbit solutions distributed over 5$\sigma$ confidence surfaces
also yield libration for at least 3 Myr, and that orbit solutions
distributed over 1$\sigma$ confidence surfaces
yield libration for up to 1 Gyr for our 1:1 resonant
object and for a subset of our 5:2 resonant objects.

Object 2001QR$_{322}$, the first discovered Neptunian Trojan,
librates about the leading Lagrange (L4) point of Neptune.
Numerical integrations of its trajectory that account for the presence
of the four giant planets reveal that libration persists
for at least 1 Gyr. Furthermore, the orbital elements of 2001QR$_{322}$
and its small libration amplitude of $24\degr$ are consistent
with the properties of Neptunian Trojans that
are stable for 4 Gyr, as described
by, e.g., Nesvorny \& Dones (2002).
It seems unlikely that the Trojan
was captured into the 1:1 MMR purely by dint of Neptune's
hypothesized migration; as Neptune encroaches upon an object,
the latter is sooner scattered onto a highly eccentric and inclined
orbit than caught into co-orbital resonance. More probably, Neptunian Trojans
pre-date the migration phase and owe their existence
to the same process that presumably gave rise to the Jovian Trojans:
trapping of planetesimals into libration about the L4/L5 points
of an accreting protoplanetary core (Marzari \& Scholl 1998; Fleming \&
Hamilton 2000). Subsequent outward radial migration
by Neptune due to the scattering of planetesimals
causes $\sim$20--82\% of Neptunian Trojans
to escape the resonance (Gomes 1998), and
decreases the eccentricities and inclinations of the remaining bound
fraction by modest amounts, $\sim$10\% at most. Taken together,
the long-term stability of 2001QR$_{322}$ (this paper; Nesvorny \&
Dones 2002) and its relative insensitivity to dramatic changes in
Neptune's orbit
lead us to regard Neptunian Trojans as dynamically pristine
compared to the rest of the Kuiper belt. The 1:1 MMR
acts as a shelter against close encounters
and dynamical excitation by resonance sweeping.
It is possible that the Trojan
has always remained confined to heliocentric distances
of 20--30 AU.

For an assumed albedo of 12--4\%, our Neptune Trojan is
130--230 km in diameter. When our DES search fields are overlaid on
model-dependent
predictions of the sky density of Neptunian Trojans constructed
by Nesvorny \& Dones (2002), we estimate that between $\sim$20
and $\sim$60 Neptune Trojans resembling 2001QR$_{322}$ librate
about Neptune's L4 and L5 points. For comparison, about $\sim$10 Jovian
Trojans exist having diameters between 100 and 200 km (Davis et al.~2003).

Our Deep Ecliptic Survey has uncovered
3 members of the 5:2 MMR. Among all resonant
KBOs, these objects possess the highest orbital eccentricities
and substantial orbital inclinations.
Their orbits cannot be a consequence of the standard
model of Neptune's migration; they cannot have originated
from low-$e$, low-$i$ orbits that underwent resonant capture
and adiabatic excitation by a migratory Neptune.
The probability of resonant capture into the 5:2 MMR under
cold initial conditions is too small compared with
the probabilities of capture into the 2:1 and 3:2 MMRs; the standard
model predicts a population ratio of $\sim$0.01 between the 5:2 and 2:1 MMRs
that is not easily reconciled with the observed ratio of $\sim$3.
Moreover, the orbital eccentricities and inclinations of 5:2 resonant
objects that are predicted by the standard model are too low compared with
their observed values.

The inability of the 5:2 MMR to capture objects on low
eccentricity orbits is reflected in the vanishingly small width
of the resonance at $e = 0$. By contrast, we have found
by numerical simulations akin to those that produced
Figure \ref{width} that the 2:1 and 3:2
MMRs both {\it widen} as $e$ decreases from 0.05 to 0. Note that
estimates by Malhotra (1996) of resonant widths do
not extend to $e < 0.05$. Murray \& Dermott (1999; see their
Figure 8.7) discuss this qualitative difference between the low-eccentricity
behavior of first-order interior MMRs and that of higher-order interior MMRs.
We reserve a more detailed theoretical exploration of the
dynamics and capture efficiencies of exterior MMRs
to a future study.

The simplest channel for populating the 5:2
MMR with objects like those that we have observed
involves adiabatically slow sweeping of that MMR
over a {\it pre-heated} Kuiper belt, i.e., one containing
a significant proportion of
initially high-$e \gtrsim 0.2$, high-$i \gtrsim 0.2$
orbits prior to the migration phase.
Capture efficiencies increase at large $e$ for
the 5:2 resonance, a reflection of the
greater width of this resonance at large $e$
and its vanishingly small width at small $e$.
The libration amplitudes predicted by resonance sweeping
over a pre-heated belt are moderate, between $16\degr$
and $145\degr$, and accord well with those observed.

Direct scattering of objects into the 5:2 and other resonances
via close encounters with Neptune can also generate the large
eccentricities and inclinations that are observed, but
generally fails to reproduce the observed libration properties.
Direct scattering yields objects that are barely bound to MMRs;
large libration amplitudes exceeding $160\degr$ are predicted
for the 5:2, 2:1, and 3:2 MMRs, in conflict with the observations.
Additional mechanisms---gravitational interactions and/or physical
collisions with bodies other than Neptune---would need
to be invoked to dampen libration amplitudes.
Levison \& Stern (1995) elaborate on a series of events
that can dampen the libration amplitude
of Pluto in the 3:2 resonance; analogous events
would need to be invoked to dampen the libration amplitudes
of members of other MMRs.
These mechanisms require the ancient Kuiper belt to be orders of magnitude
more populous than it is today, a prospect that by itself
does not appear unreasonable, given the requirements of planet
formation models (Kenyon 2002), and ongoing dynamical (Holman \& Wisdom
1993; Duncan, Levison, \& Budd 1995) and collisional erosion of the belt.

Nonetheless, we feel that Occam's razor, and the physical plausibility and
seeming inevitability of planetary
migration driven by planetesimal scattering (Fernandez \& Ip 1984;
Hahn \& Malhotra 1999), would seem to disfavor resonance population
mechanisms that do not invoke migration. Perhaps the principal
objection to this mechanism lies in the possibility that
Neptune's migration was insufficiently smooth to resonantly
capture objects; we discuss quantitatively this possibility
in the Appendix. Chiang \& Jordan (2002)
offer a more objective test of the migration hypothesis;
for sufficiently fast migration rates, the number of
2:1 resonant KBOs having libration centers
$\langle \phi_{2:1} \rangle \approx 270\degr$
should exceed those having $\langle \phi_{2:1} \rangle \approx 90\degr$
by factors of $\sim$3. An asymmetry in libration center populations
translates directly into an asymmetry in the sky density
of Twotinos about the Sun-Neptune line.
By contrast, if the 2:1 resonance were populated by direct
scattering, the libration centers would presumably be equally populated.
At present, the number (6) of known Twotinos is too small
to permit the drawing of firm conclusions.

In summary, it is most straightforward to reproduce the observed pattern
of resonance occupation in the Kuiper belt by presupposing
both initially cold orbits (to populate resonances such
as the 3:2 and 2:1 MMRs) and initially hot orbits (to populate
resonances such as the 5:2 MMR) prior to Neptune's migration.

What might have heated the primordial Kuiper belt prior
to resonance sweeping? Models of planetesimal formation
predict eccentricities and inclinations of less than $\sim$0.05
(see, e.g., Kenyon \& Luu 1999; Lissauer 1993; Kokubo \& Ida 1992),
values below what are required to capture KBOs into the 5:2 resonance
with the relative efficiencies observed.
Thommes, Duncan, \& Levison (1999, 2002) propose that
the $\sim$10 $M_{\oplus}$ embryonic cores of Neptune and
Uranus were scattered into the ancient belt and heated KBOs
by dynamical friction. A possible problem with this scenario
is that Neptune's orbital history may be too violent to generate
a significant Trojan population.

By contrast, we believe there exists another pre-heating mechanism
that is more natural: scattering of KBOs to large $e$ and large $i$ by Neptune
as that planet migrated outwards (Gomes 2002). The formation
of a primitive scattered disk that is later swept over by
mean-motion resonances seems a natural consequence of planetary
migration driven by planetesimal scattering. What requires
further elucidation is how the perihelia of the scattered
objects are raised so as to avoid further scatterings by Neptune.

\acknowledgements
We are indebted to Dr.~Jana Pittichova for donating her
telescope time to secure
astrometric observations of our
Neptune Trojan that helped to solidify its dynamical
identity.
EIC acknowledges support from
National Science Foundation Planetary Astronomy Grant
AST-0205892, Hubble Space Telescope Theory Grant HST-AR-09514.01-A,
and a Faculty Research Grant awarded by the University
of California at Berkeley. MWB, RLM, and LHW are supported in
part by NASA grants NAG5-8990 and NAG5-11058. Research
by JLE and SDK is supported, in part, by NASA grant NAG5-10444.
DET is supported by grants from the American Astronomical Society
and the Space Telescope Science Institute.
KJM is supported by NASA grant NAG5-4495.
We thank Kelly Clancy and Mark Krumholz for assistance,
and David Nesvorny for a thoughtful referee's report that
significantly improved this paper. The NOAO observing facilities
used in the Deep Ecliptic Survey are
supported by the National Science Foundation.

\begin{appendix}
\section{Breakdown of Smooth Migration}

Here we crudely estimate the critical sizes
of Neptune-encountering planetesimals above
which our assumption of smooth migration would be invalid.
We imagine that at each instant during Neptune's migration, Neptune's
annulus of influence---$r_H \sim (m_N/3m_{\odot})^{1/3} a_N$ in
extent, where $m_N$ and $m_{\odot}$ are the mass of Neptune
and of the Sun, respectively, and $a_N$ is the semi-major
axis of Neptune---contains
$N$ planetesimals each having $\Delta m$ mass.
A typical Poisson fluctuation in the number of planetesimals
is $\sqrt{N}$. It is this random fluctuation that generates
a change of random sign in Neptune's semi-major axis, by an amount
of order $\delta a \sim \sqrt{N} (\Delta m/m_N) a_N$ after
all the planetesimals within the annulus of influence are scattered
away. The duration of encounter between each planetesimal and Neptune
is of order Neptune's orbital period, $P_N$. Then the magnitude
of the random component of Neptune's migration rate is of order
$\delta a / P_N$.

For the migration to be smooth,

\begin{equation}
\delta a / P_N < \dot{a}_{mean} \, ,
\label{condi}
\end{equation}

\ni where $\dot{a}_{mean}$ is the mean (smooth) migration rate
that arises from the mean difference in Neptune-encountering
fluxes having high specific angular momentum and fluxes having
low specific angular momentum.
In our simulations, $\dot{a}_{mean} = \Delta a_N \exp(-t/\tau) / \tau$,
where $\Delta a_N = 7 \AU$ and $\tau = 10^7\yr$.

We estimate the surface mass density of planetesimals within
the annulus of influence to be the surface mass density of
planetesimals throughout the disk. Hahn \& Malhotra (1999)
find in their numerical simulations of planetary migration
(using effective particle sizes too large to engender smooth
migration) that $m_{disk} \sim 50 M_{\oplus}$ of material
must be interspersed between
$a = 10\AU$ and $a_{disk} = 50\AU$ to drive Neptune's orbit outward
by $\Delta a_N \approx 7 \AU$.
Then our order-of-magnitude
estimate for the surface mass density everywhere is
$N\Delta m/ \pi r_s^2 \sim m_{disk}/\pi a_{disk}^2$.
We solve this equation for $N$, and insert
into condition (\ref{condi}) to find

\begin{equation}
\frac{\Delta m}{m_N} < \left( \frac{a_{disk}}{a_N} \right)^2 \left(
\frac{\Delta a_N}{r_H} \right)^2 \left( \frac{P_N}{\tau} \right)^2 \left(
\frac{m_N}{m_{disk}} \right) \exp(-2t/\tau) \,.
\end{equation}

\ni For $m_N = 17 M_{\oplus}$, $a_N = 25 \AU$, $r_H = 0.6 \AU$,
$P_N = 130 \yr$, $t = \tau = 10^7\yr$, and other parameters
as listed above,
we find that $\Delta m / m_N < 4 \times 10^{-9}$
for the migration to be smooth. Spheres of density $2 \gm \cm^{-3}$
having radii less than 40 km would suffice. If Neptune's migration
instead occurred over timescales of $\tau = 3\times 10^6\yr$,
the critical radius grows to 80 km. Kenyon (2002) calculates
that 90\% of the mass of the primordial Kuiper belt was contained in
bodies having radii of 0.1--10 km at heliocentric distances
of 40--50 AU. It is not clear, however, how these accretion calculations
should be modified at distances of 20--30 AU where Neptune
resided. An inopportune encounter between Neptune
and a single massive planetesimal might have caused the former to lose
whatever retinue of resonant objects it had accumulated previously
by smooth migration.
\end{appendix}

\newpage
\begin{deluxetable}{lccccccc}
\tabletypesize{\scriptsize}
\tablewidth{0pt}
\tablecaption{Orbital Elements\tablenotemark{a}~~of DES 5:2 and 1:1 Resonant
KBOs}
\tablehead{
\colhead{Name} &
\colhead{Resonance} & \colhead{$a$ (AU)} &
\colhead{$e$} & \colhead{$i$ (deg)} & \colhead{$\Omega$ (deg)} &
\colhead{$\omega$ (deg)} & \colhead{$M$ (deg)}
}
\startdata
1998WA$_{31}$ & 5:2 & 55.73 & 0.432 & 9.43 & 20.7 & 310.7 & 28.2 \\
(38084) 1999HB$_{12}$ & 5:2 & 55.10 & 0.409 & 13.17 & 166.5 & 66.7 & 343.1 \\
2001KC$_{77}$ & 5:2 & 54.67 & 0.352 & 12.9 & 57.8 & 181.8 & 358.4 \\
2001QR$_{322}$ & 1:1 & 30.39 & 0.028 & 1.32 & 151.6 & 236.2 & 327.3 \\
\enddata
\tablenotetext{a}{Osculating, heliocentric elements referred to the J2000
ecliptic plane, evaluated at epoch 2451545.0 JD. Elements shown here are
best-fit values; for a discussion of uncertainties, see \S\ref{clasproc} and
Elliot et al.~(2003). The angles $\Omega$, $\omega$, and $M$ are the longitude
of ascending node, the argument of perihelion, and the mean anomaly,
respectively.}
\end{deluxetable}
\newpage
\begin{deluxetable}{lccccccc}
\tabletypesize{\scriptsize}
\tablewidth{0pt}
\tablecaption{Designations of DES and Non-DES Resonant KBOs}
\tablehead{
\colhead{Resonance} &
\colhead{Name}
}
\startdata
1:1 & 2001QR$_{322}^{\ast}$ \\
5:4 & 1999CP$_{133}$ \\
4:3 & 1998UU$_{43}^{\ast}$, 2000CQ$_{104}^{\ast}$, (15836) 1995DA$_2$ \\
3:2 & (28978) Ixion$^{\ast}$, 1998UR$_{43}^{\ast}$, 1998US$_{43}^{\ast}$,
1998WS$_{31}^{\ast}$, 1998WU$_{31}^{\ast}$, 1998WV$_{31}^{\ast}$, \\
    & 1998WW$_{24}^{\ast}$, 1998WZ$_{31}^{\ast}$, 2000CK$_{105}^{\ast}$,
2001KY$_{76}^{\ast}$, 2001KB$_{77}^{\ast}$, 2001KD$_{77}^{\ast}$, \\
    & 2001KQ$_{77}^{\ast}$, 2001QF$_{298}^{\ast}$, 2001QG$_{298}^{\ast}$,
2001RU$_{143}^{\ast}$, 2001RX$_{143}^{\ast}$, (15788) 1993SB, \\
    & (15789) 1993SC, (15810) 1994JR$_1$, (15820) 1994TB, (15875)
1996TP$_{66}$, (19299) 1996SZ$_4$, \\
    & (20108) 1995QZ$_9$, (24952) 1997QJ$_4$, (32929) 1995QY$_9$, (33340)
1998VG$_{44}$, (38628) 2000EB$_{173}$, \\
    & (47171) 1999TC$_{36}$, (47932) 2000GN$_{171}$, 1993RO, 1995HM$_5$,
1996RR$_{20}$, 1996TQ$_{66}$, 1998HH$_{151}$, \\
    & 1998HK$_{151}$, 1998HQ$_{151}$, 1999CE$_{119}$, 1999CM$_{158}$,
1999TR$_{11}$, 2000FV$_{53}$, \\
    & 2000GE$_{147}$, 2001FL$_{194}$, 2001VN$_{71}$, 2001YJ$_{140}$,
2002VE$_{95}$, 2001FU$_{172}$ \\
5:3 & (15809) 1994JS, 1999CX$_{131}$, 2001XP$_{254}$ \\
7:4 & 2000OP$_{67}^{\ast}$, 2001KP$_{77}^{\ast}$, 1999KR$_{18}$,
1999RH$_{215}$, 2000FX$_{53}$, 2000OY$_{51}$ \\
2:1 & 2000QL$_{251}^{\ast}$, (20161) 1996TR$_{66}$, (26308) 1998SM$_{165}, $ \\
    & 1997SZ$_{10}$, 1999RB$_{216}$, 2000JG$_{81}$ \\
7:3 & 1999CV$_{118}$ \\
5:2 & (38084) 1999HB$_{12}^{\ast}$, 1998WA$_{31}^{\ast}$, 2001KC$_{77}^{\ast}$,
\\
    & (26375) 1999DE$_9$, 2000FE$_8$, 2000SR$_{331}$, 2002TC$_{302}$ \\
\enddata
\tablenotetext{a}{Objects discovered by DES are denoted by an asterisk.}
\end{deluxetable}

\end{document}